\documentclass[12pt]{iopart}
\usepackage{graphicx}
\usepackage{dcolumn}
\usepackage{bm}
\usepackage{amssymb}

\begin{document}
\title{Screening of heterogeneous surfaces: charge renormalization of Janus particles}
\author{N. Boon$^1$, E. Carvajal Gallardo$^1$, S. Zheng$^1$, E. Eggen$^1$, M. Dijkstra$^2$ and R. van Roij$^1$,}
 \address{$^1$ Institute for Theoretical Physics, Utrecht University, Leuvenlaan 4, 3584 CE Utrecht, The Netherlands}

\address{$^2$ Soft Condensed Matter, Debye Institute for Nanomaterials Science, Utrecht University,
Princetonplein 5, 3584 CC Utrecht, The Netherlands}

\begin{abstract}
Nonlinear ionic screening theory for {\em heterogeneously} charged spheres is developed in terms of a
mode-decomposition of the surface charge. A far-field analysis of the resulting electrostatic potential leads to a
natural generalization of charge renormalization from purely monopolar to dipolar, quadrupolar, etc., including
`mode-couplings'. Our novel scheme is generally applicable to large classes of surface heterogeneities, and is
explicitly applied here to Janus spheres with differently charged upper and lower hemispheres, revealing strong
renormalization effects for all multipoles.

\end{abstract}
\maketitle

\section{Introduction \label{introduction}}
The past few years have seen an explosion of newly synthesized colloidal (nano-)particles that are {\em not}
spherically symmetric, either by shape (e.g. dumbbells, snowmen, cubes) or by surface pattern (patches, stripes)
\cite{glotzer}.
A broken rotational symmetry also occurs when particles are adsorbed to air-water or oil-water interfaces
\cite{Binks1}, as in for example, colloidal monolayers \cite{pieranski,frydel}, Pickering emulsions \cite{pickering}, or bijels
\cite{bijels}.
Moreover, recent atomic force microscopy studies have shown that even supposedly homogeneous colloidal surfaces
can actually be heterogeneous on length scales as large as 100 nm \cite{tong}, while atomic corrugations and
facets render any nm-sized particle strictly heterogeneous. An important consequence of surface heterogeneity is
anisotropy of the mutual effective forces, which directly affects the self-assembly process of the (nano)colloids
into large-scale structures, for instance into ill-understood linear chains \cite{frydel,tong,strings}.

A fundamental problem is thus to establish relationships between shape and surface heterogeneity on the one hand
and effective interactions and large-scale self-assembly structures on the other \cite{glotzer}. Apart from
specific forces (e.g., hydrophobic, van der Waals) the effective interactions between dispersed particles often
involve a strong generic electrostatic component, which is well-described, for homogeneously charged objects, by
linear-screening theory {\em provided} renormalized charges instead of bare charges are used \cite{alexander}. Renormalization of heterogeneously distributed surface charge is an open problem, for which we develop a systematic theory in this paper. We go beyond recent linear screening treatments \cite{linear} and formulate a new and efficient framework for computing {\em nonlinear} ionic screening effects of heterogeneously charged
spheres dispersed in a 1:1 electrolyte. Our theory generalizes Alexander's notion of ion-condensation induced
charge-renormalization \cite{alexander} to include not only the monopole but also the dipole, quadrupole, etc., as
well as their nonlinear couplings. These multipole modes can be important if one wishes to calculate the electrostatic force between particles, which was already shown for clay platelets \cite{clay}. Our scheme is versatile and can be applied to essentially any type of charge
heterogeneity. We focus on applications to Janus spheres composed of two differently charged hemispheres
\cite{janus}.

We consider an index-matched suspension of $N$ colloidal spheres of radius $a$ in a bulk solvent of dielectric
constant $\epsilon$ and volume $V$ at temperature $T$. The solvent also contains point-like monovalent cations
(charge $+e$) and anions (charge $-e$) at fugacity $\rho_s$. Here $e$ is the elementary charge. A relatively
simple treatment of this many-body problem is the cell model \cite{alexander}, in which a single colloid is
considered in the center of a spherical cell of radius $R$ and volume $(4\pi/3)R^3\equiv V/N$. We denote the
surface charge density of this central colloidal particle by $e\sigma(\theta,\varphi)$, where $\theta$ and
$\varphi$ are the standard polar and azimuthal angle, respectively, with respect to a laboratory frame. Within a
mean-field approximation, the concentration profiles of the cations and anions can be written as Boltzmann
distributions $\rho_{\pm}({\bf r})=\rho_s\exp[\mp\Phi({\bf r})]$, where $k_BT\Phi({\bf r})/e$ is the electrostatic
potential at ${\bf r}=(r,\theta,\varphi)$, with $k_B$ the Boltzmann constant and $r=|{\bf r}|$. Note that
$\Phi({\bf r})=0$ in the salt reservoir, and that $\rho_{\pm}({\bf r})=0$ for $r<a$ due to hard-core exclusion.
The potential must satisfy the Poisson equation $\nabla^2\Phi({\bf r})=-4\pi\lambda_B(\rho_+({\bf r})-\rho_-({\bf
r}))$, where we defined the Bjerrum length $\lambda_B=e^2/\epsilon k_BT$.  Combining the Poisson and Boltzmann
equations gives
\begin{eqnarray}
\!\!\!\!\Phi''({\bf r})+\frac{2\Phi'({\bf r})}{r}-\frac{{\cal L}^2\Phi({\bf
r})}{r^2}&=&\left\{\begin{array}{ll}0&r<a;\\\kappa^2\sinh\Phi({\bf r})&r>a,\end{array}\right.\label{PB}
\end{eqnarray}
where a prime denotes a radial derivative,  ${\cal
L}^2=-[(\sin\theta)^{-1}\partial_{\theta}\sin\theta\partial_{\theta} +
(\sin\theta)^{-2}\partial^2_{\varphi\varphi}]$ the {\em angular momentum operator},  and
$\kappa^{-1}=(8\pi\lambda_B\rho_s)^{-1/2}$ the screening length. On the colloidal surface, $r=a$, Gauss' law
imposes the boundary condition (BC)
\begin{eqnarray}
 \lim_{r \downarrow a}\Phi'(r,\theta,\varphi)=\lim_{r \uparrow a} \Phi'(r,\theta,\varphi)
-4\pi\lambda_B\sigma(\theta,\varphi).\label{gauss}
\end{eqnarray}
Electro-neutrality of the cell imposes $\int d\varphi d\theta\sin\theta \Phi'(R,\theta,\varphi)=0$ at $r=R$, which is a
sufficiently stringent BC to close the system of equations in the spherically symmetric case. Now, however, an
additional BC is to be specified for the angular dependence at $r=R$, depending on the environment of the cell. For now, we assume an environment that is characterized by `isotropic' boundary conditions,
\begin{eqnarray}
\partial_{\theta}\Phi(R,\theta,\varphi)=\partial_{\varphi}\Phi(R,\theta,\varphi)=0. \label{BCiso}
\end{eqnarray}
We will discuss this choice, and its consequences, in section \ref{discussion}.

\section{Theory \label{method}}
For a given $\sigma(\theta,\varphi)$ one can solve \eref{PB} with BC's for $\Phi({\bf r})$, for example, numerically on a discrete $(r,\theta,\varphi)$ grid. The approach we take, however, avoids a cumbersome 3-dimensional grid in
favor of a systematic expansion of the angular dependence in spherical harmonics. For notational convenience and
illustration purposes we restrict attention here to $\varphi$-independent cases where the expansion involves only
Legendre polynomials $P_{\ell}(x)$ with $x=\cos\theta$.

The first step in this analysis is the decomposition of the colloidal surface charge into surface multipoles
$\sigma_{\ell}=\frac{2\ell+1}{2}\int_{-1}^1 dx\sigma(x)P_{\ell}(x)$, such that
$\sigma(x)=\sum_{\ell=0}^{\infty}\sigma_{\ell}P_{\ell}(x)$. Similarly we decompose
$\Phi(r,x)=\sum_{\ell=0}^{\infty}\Phi_{\ell}(r)P_{\ell}(x)$. With $\Phi_{\ell}(r) = (r/a)^{\ell}\Phi_{\ell}(a)$
the regular solution to (\ref{PB}) for $r\in[0,a]$, the BC's for $r\in\{a,R\}$,
\begin{eqnarray}
\Phi_{\ell}'(a)&=&\frac{\ell}{a} \Phi_{\ell}(a) - 4\pi\lambda_B\sigma_{\ell}\,\,\,\,\,\,\,\,(\ell\geq 0);\label{gauss2}\\
 \Phi_0'(R)&=&0 \,\,\,\,\,\mbox{and}\,\,\,\,\,\, \Phi_{\ell}(R)=0 \,\,\,\,\,(\ell\geq 1),  \label{BCiso2}
\end{eqnarray}
conveniently decouple for the different $\ell$'s. By contrast, the nonlinear $\sinh$ term in (\ref{PB}) induces
`mode-coupling' between {\em all} Legendre components $\Phi_{\ell}(r)$ -- not to be confused with mode-couplings in dynamical slowing down. This coupling is obviously unpractical for a numerical treatment.

The second step of our analysis resolves this mode-coupling problem by `ordering' the modes systematically. We
introduce a dimensionless `switching' parameter $A$, and consider the auxiliary distribution
$\sigma^{(A)}(x)=\sum_{\ell=0}^{\infty}A^{\ell}\sigma_{\ell}P_{\ell}(x)$, such that $A=0$ describes a homogeneous
distribution and $A=1$ the heterogeneous one of interest. We also define the corresponding
auxiliary potential $\Phi^{(A)}(r,x)=\sum_{n=0}^L A^n\phi_n(r,x)$, where $L$ sets the order of the truncation and
where the expansion coefficients can themselves be expanded as
$\phi_{n}(r,x)=\sum_{\ell=0}^nf_{n\ell}(r)P_{\ell}(x)$.
The functions $f_{n\ell}(r)$ are independent of $A$ and will be calculated numerically below for the cases of
interest $n\geq \ell$ (which is assumed implicitly from now on). Since the problem is invariant under the
simultaneous transformation $A\rightarrow -A$ and $x\rightarrow -x$ one checks that $f_{n\ell}(r)=0$ for $n+\ell$
odd, that is, we only consider $\ell$ and $n$ both even or both odd.

Replacing $\sigma(x)$ by $\sigma^{(A)}(x)$ and $\Phi(r,x)$ by $\Phi^{(A)}(r,x)$, inserting the corresponding
expansions into the BC's, and equating all orders of $A$ yields at $r\downarrow a$ and $r=R$
\begin{eqnarray}
f_{n\ell}'(a)&=& \frac{\ell}{a} f_{n\ell}(a)  -4\pi\lambda_B\sigma_{\ell}\delta_{n\ell}; \label{gauss3}\\
f_{n\ell}'(R)&=&0\, \,\,(\ell=0)\,\, \mbox{and}\,\,f_{n\ell}(R)=0 \,\,\,(\ell\geq 1),\label{BCiso3}
\end{eqnarray}
where $\delta_{n\ell}$ is the Kronecker-delta. When the same replacement and expansion procedure is applied to the
PB equation (\ref{PB}), one finds upon expanding the argument of the $\sinh$ term with respect to $A$ a hierarchy
of second-order differential equations for $f_{n\ell}(r)$, with a structure that allows for an order-by-order
sequential solution. For $n=\ell=0$ we obtain  for $r\in[a,R]$ the spherically symmetric nonlinear PB equation in
the cell, $f_{00}''(r)+\frac{2f_{00}'(r)}{r}=\kappa^2\sinh f_{00}(r)$,
which we solve explicitly with the BC's given in (\ref{gauss3}) and (\ref{BCiso3}) on a radial grid. We thus
consider $f_{00}(r)$ as a known function from now on. For $n\geq1$ we obtain
\begin{equation}
f_{n\ell}''(r)+\frac{2f_{n\ell}'(r)}{r}-\Big(\frac{\ell(\ell+1)}{r^2}+\kappa^2\cosh
f_{00}(r)\Big)f_{n\ell}(r)=\kappa^2 S_{n\ell}(r), \label{fnl}
\end{equation}
where $S_{n\ell}(r)$ acts as a source term of the form $S_{11}=0$, $S_{20}=\frac{1}{6}f_{11}^2\sinh f_{00}$,
$S_{22}=2S_{20}$, $S_{31}=f_{11}(f_{20}+\frac{2}{5}f_{22})\sinh f_{00} + \frac{1}{10}f_{11}^3\cosh f_{00}$,
$S_{33}=\frac{3}{5}f_{11}f_{22}\sinh f_{00} + \frac{1}{15}f_{11}^3\cosh f_{00}$, and explicit expressions for
higher-order terms can be generated straightforwardly. The key observation is that $S_{n\ell}$ {\em only} depends
on $f_{n'\ell'}$'s with $n'<n$, that is, a hierarchy of terms follows spontaneously. Thus \eref{fnl} with the BC's (\ref{gauss3}) and (\ref{BCiso3}) can be solved for $n=\ell=1$, which in turn determines $S_{20}(r)$ and
$S_{22}(r)$ such that $f_{20}(r)$ and $f_{22}(r)$ can be solved, etc. The nonlinear mode-coupling, represented
explicitly by $\cosh f_{00}(r)$ and $S_{n\ell}(r)$, renders the linear equation \eref{fnl} highly nontrivial, yet
numerical solution on a radial grid $r\in[a,R]$ is straightforward. With $f_{n\ell}(r)$ determined for $L\geq
n\geq\ell\geq0$ for some cut-off $L$, we can set $A=1$ to explicitly construct the potential of interest
$\Phi(r,x)=\sum_{\ell=0}^L\Phi_{\ell}(r)P_{\ell}(x)$ with $\Phi_{\ell}(r)=\sum_{n=\ell}^Lf_{n\ell}(r)$.

\begin{figure}[!ht]
\centering
\includegraphics[width = 8cm]{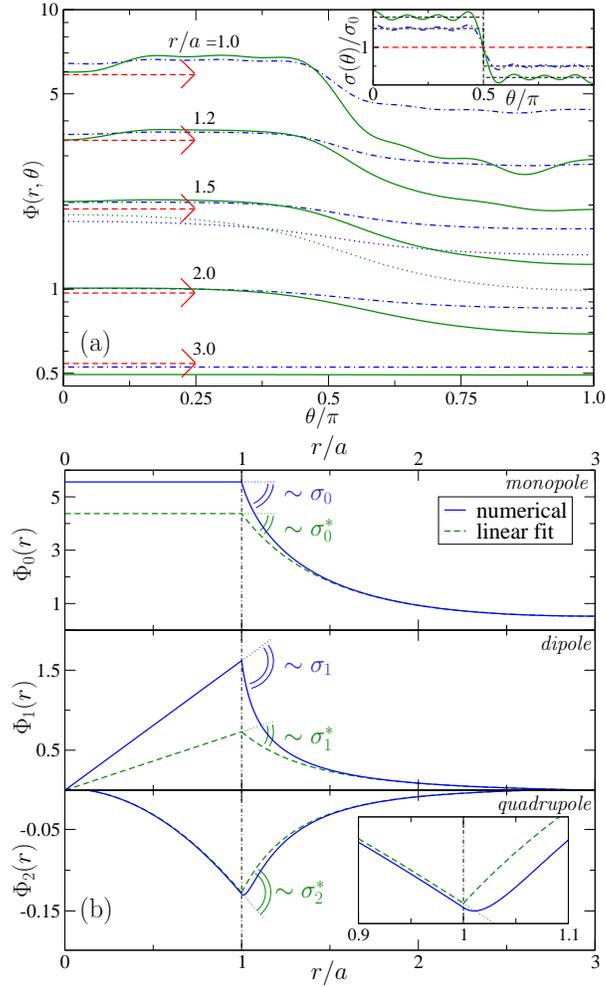}
\caption{(a) Angular dependence of the potential (at several distances $r$) and the surface charge (inset) of a
Janus sphere of radius $a$, for $\kappa a =1$, $Z_0\lambda_B/a=22.5$,  $R/a=3$, and $L=13$, for charge
heterogeneities (see text) $g=0$ (arrows), $g=0.5$ (dot-dashed), and $g=0.8$ (solid curves). The dotted curves for
$r/a=1.5$ and $g\in\{0.5,0.8\}$ stem from a Yukawa-segment model (see text). (b) Radial dependence of the monopole
($\ell=0$), dipole ($\ell=1$), and quadrupole ($\ell=2$) potentials $\Phi_{\ell}(r)$ (solid lines), and their
far-field linear-screening approximations (dashed curves), for the parameters of (a) and $g=0.5$. The angles
defined at $r=a$ relate to the bare and renormalized modes of the surface charge (see text). } \label{fig:FIGURE1}
\end{figure}

\section{Results \label{results}}
The theory developed so far is directly applicable to any uniaxial charge distribution, while generalizations to
azimuthal dependencies and nonspherical shapes are feasible. In this paper we illustrate our scheme for the
prototype heterogeneous charge distribtion of Janus spheres characterized by surface charge-densities $\sigma_N$
and $\sigma_S$ on the northern $(x>0)$ and southern ($x<0$) hemisphere, respectively \cite{janus,janus2}. The
non-vanishing modes are thus $\sigma_0=(\sigma_N+\sigma_S)/2$ and $\sigma_{\ell}=g\sigma_0
(2\ell+1)(-1)^{\small\frac{\ell-1}{2}}\frac{(\ell-2)!!}{(\ell+1)!!}$ for $\ell$ odd. Here we defined the
dimensionless heterogeneity parameter
\begin{equation}
g=\frac{\sigma_N-\sigma_S}{\sigma_N+\sigma_S} \label{heterogeneity},
\end{equation}
which together with the total charge $Z_0=4\pi a^2\sigma_0$ fully characterizes the distribution.  Below we set $\kappa a=1$, $R/a=3$, and
$L=13$ throughout unless stated otherwise, and we identify $Z_0\lambda_B/a$ as the only other relevant
dimensionless combination. For $Z_0\lambda_B/a=22.5$ and $g\in\{0, 0.5, 0.8\}$ figure 1(a) shows the
$\theta$-dependence of $\sigma$ (inset) and $\Phi$ for several $r$, revealing isotropy (as expected) for a
homogeneous surface charge ($g=0$, arrows) and strong anisotropy for the heterogeneous cases $g=0.5$
$(\sigma_N=3\sigma_S$, dot-dashed) and $g=0.8$ ($\sigma_N=9\sigma_S$, solid lines); the (small) oscillations with
$\theta$ at $r=a$ are numerical artefacts due to the truncation at $L=13$.  The $\theta$-dependence of the
potential weakens, as expected, for increasing distances $r$. Figure 1(b) shows $\Phi_{\ell}(r)$ for $\ell=0,1,2$ and
$g=0.5$. Interestingly, the modes with $\ell=1,2$ have a non-vanishing electric field in the interior of the
particle. The overall magnitude and spatial variation of $\Phi$ in figure 1 show the need for nonlinear screening
theory. Nevertheless, in analogy to the spherically symmetric case \cite{alexander} one can describe the far-field
potential ($r\simeq R$) and hence the colloidal interactions in terms of {\em linear} screening theory (dashed
curves in figure 1(b)) with a renormalized surface charge distribution $\sigma^*(x)\equiv
\sum_{\ell}\sigma^*_{\ell}P_{\ell}(x)$ that we will calculate below.

In the far-field $r\simeq R$ we treat the deviation of $\Phi(r,x)$ from its angular average
$\Phi_0(R)\equiv\chi_0$ at $r=R$ as a small expansion parameter, such that (\ref{PB}) for $r>a$ can be linearized as
$\nabla^2\Phi(r,x)\simeq\bar{\kappa}^2[\tanh\chi_0 + (\Phi(r,x)-\chi_0)]$ with
$\bar{\kappa}^2=\kappa^2\cosh\chi_0$. The uniaxial solutions to this linear PB (LPB) equation read
$\Phi(r,x)\simeq\chi_0-\tanh\chi_0+\sum_{\ell=0}^{\infty}[a_{\ell}i_{\ell}(\bar{\kappa}r) +
b_{\ell}k_{\ell}(\bar{\kappa}r)]P_{\ell}(x)$
where $i_{\ell}$ and $k_{\ell}$ are modified spherical Bessel functions. The coefficients $a_{\ell}$ and
$b_{\ell}$ are integration constants that we fix by matching the LPB-solution at $r=R$, for each $\ell$, to
$\Phi_{\ell}(R) \equiv \chi_{\ell}$ and $\Phi'_{\ell}(R) \equiv \chi_{\ell}'$ of the {\em nonlinear} problem. This leads for every $\ell$ to the linear two by two problem
\begin{eqnarray}
\chi_{\ell}&=&(\chi_0-\tanh\chi_o)\delta_{\ell 0} + a_{\ell}i_{\ell}(\bar{\kappa}R) +
b_{\ell}k_{\ell}(\bar{\kappa}R);
\nonumber\\
\chi_{\ell}'&=&\bar{\kappa}\Big(a_{\ell}i'_{\ell}(\bar{\kappa}R) + b_{\ell}k'_{\ell}(\bar{\kappa}R)\Big),
\end{eqnarray}
which results in explicit expressions for $a_{\ell}$ and $b_{\ell}$ given by
\begin{eqnarray}
a_{\ell} &=& \frac{\nu_{\ell} \left(k_{\ell+1}(\bar{\kappa}R) - \frac{\ell}{\bar{\kappa}R}k_{\ell}(\bar{\kappa}R)\right)+\chi_{\ell}'k_{\ell}(\bar{\kappa}R) }
{i_{\ell}(\bar{\kappa}R) k_{\ell+1}(\bar{\kappa}R)+i_{\ell+1}(\bar{\kappa}R)k_{\ell}(\bar{\kappa}R)};\\
b_{\ell} &=& \frac{\nu_{\ell}\left(i_{\ell+1}(\bar{\kappa}R) + \frac{\ell}{\bar{\kappa}R} i_{\ell}(\bar{\kappa}R)\right) - \chi_{\ell}' i_{\ell}(\bar{\kappa}R)}
{i_{\ell}(\bar{\kappa}R) k_{\ell+1}(\bar{\kappa}R)+i_{\ell+1}(\bar{\kappa}R) k_{\ell}(\bar{\kappa}R)},
\end{eqnarray}
where $\nu_{\ell} = \chi_{\ell}-(\chi_0-\tanh\chi_o)\delta_{\ell 0}$. The dashed curves in figure 1(b) are the result of such a far-field fit.  With $a_{\ell}$ and $b_{\ell}$ explicitly known, one can extrapolate the LPB solution to $r=a$ to yield, with \eref{gauss2} and standard Bessel function relations, the renormalized multipoles
\begin{eqnarray}
\label{sigmastar} \sigma_{\ell}^*
&=&- \frac{\bar{\kappa}}{4\pi\lambda_B}\Big(a_{\ell}i_{\ell+1}(\bar{\kappa}a) -
b_{\ell}k_{\ell+1}(\bar{\kappa}a)\Big).
\end{eqnarray}
This expression is the multipole generalization of the well-known charge renormalization \cite{alexander}.

\begin{figure}[!ht]
\centering
\includegraphics[width = 8cm]{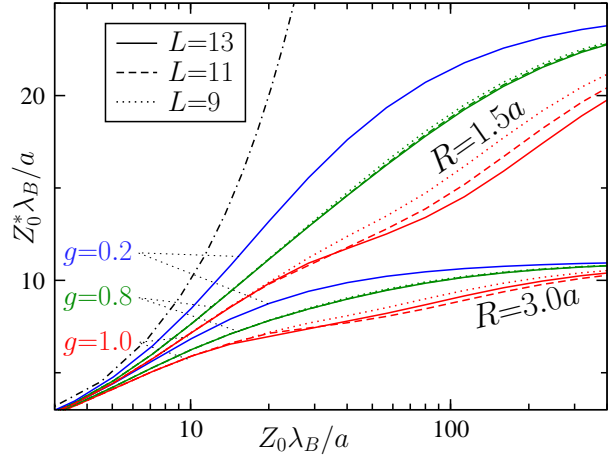}
\caption{Renormalized (scaled) monopole $Z_0^*$ for several $g$, $L$ (see text) of a Janus sphere as a function of the total charge $Z_0\lambda_B/a$. The graph includes data for cell radii $R/a = 1.5$ and $R/a = 3.0$. The dot-dashed curve denotes $Z_{0}^*=Z_{0}$. } \label{fig:FIGURE2}
\end{figure}

\begin{figure}[!ht]
\centering
\includegraphics[width = 8cm]{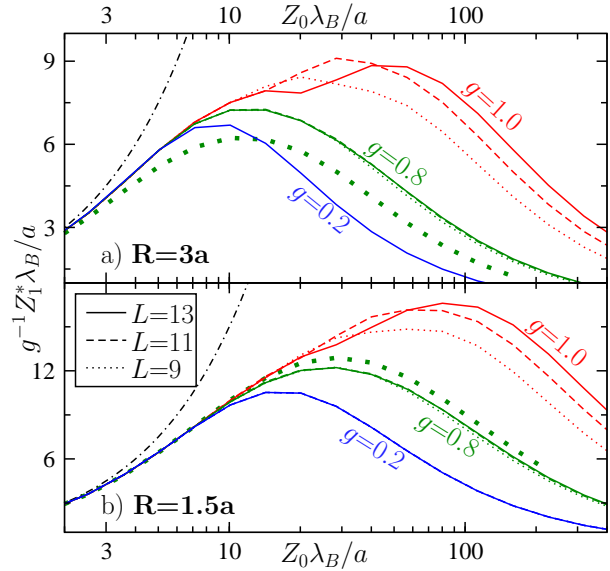}
\caption{Renormalized (scaled) dipole $Z_1^*$ for several $g$, $L$ (see text) of a Janus sphere as a function of the total charge $Z_0\lambda_B/a$. The upper and the lower graph represent cell sizes $R/a = 3.0$ and $R/a = 1.5$ respectively. The thick dotted lines denote the (scaled) difference of the renormalized northern and southern charge presumed distributed homogeneously (see text) for $g=0.8$. The dot-dashed curves in both graphs denote $Z_{1}^*=Z_{1}$. } \label{fig:FIGURE3}
\end{figure}

\begin{figure}[!ht]
\centering
\includegraphics[width = 8cm]{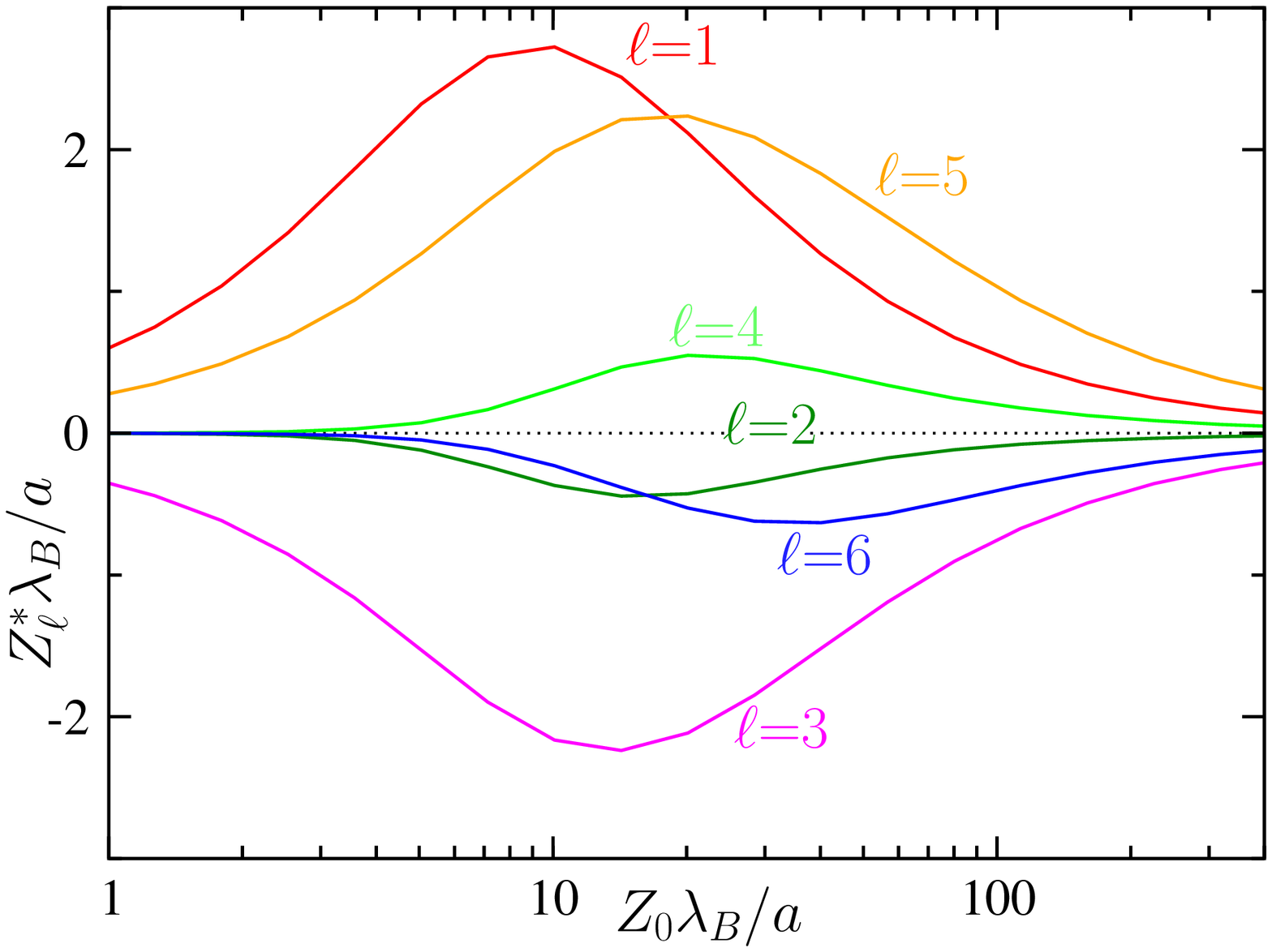}
\caption{Renormalized (scaled) multipoles $Z_{\ell}^*$  of a Janus sphere as a function of the total charge $Z_0\lambda_B/a$. The parameters here are $g=0.4$ and $L=13$.}
\label{fig:FIGURE4}
\end{figure}

A first illustration of multipole renormalization is shown by the angles in figure 1(b), which represent
slope discontinuities at $r=a$ which are proportional to $\sigma_{\ell}$ (nonlinear theory)  and $\sigma^*_{\ell}$
(far-field fit). For $\ell=0,1$ we see $\sigma_{\ell}^*<\sigma_{\ell}$, which means that the effective charge and dipole are renormalized downward. Interestingly, however, for $\ell=2$ the inset in the lowest panel reveals upward
renormalization since $\sigma_2^*\neq 0$ while $\sigma_2=0$, that is, the Janus particle has a mode-coupling induced
far-field quadrupole signature. Multipole renormalization is quantified further in figures 2--4, where (scaled)
renormalized multipoles $Z_{\ell}^*=4\pi a^2\sigma_{\ell}^*$ are shown for the monopole $\ell=0$ in figure 2, the
dipole $\ell=1$ in figure 3 (scaled with $g$), and the higher-order multipoles $\ell=1,\cdots,6$ in figure 4. All multipoles are shown as a function of
$Z_0\lambda_B/a$. Furthermore, in figure 2 and 3 we chose several truncation levels $L$ and heterogeneities $g$, and also picked
two cell radii $R$. The dot-dashed curves denote the linear limit $Z_{\ell}^*=Z_{\ell}$. Figure 2 and 3 show that all curves for $L=13$ superimpose on those of $L=9,11$ for all $Z_0$ and $g\lesssim 0.8$,
indicative of excellent convergence in this parameter regime; for $g=1$ the convergence deteriorates for
$Z_0\lambda_B/a\gtrsim 15$. The $R$ dependence in figure 2 and 3 shows the strongest renormalization in the largest cell, not
unlike the homogeneous-charge case \cite{alexander}. Interestingly, in the nonlinear regime $Z_0\lambda_B/a\gtrsim
10$ figure 2 shows a mode-coupling induced {\em reduction} of $Z_0^*$ by 10's of percents when $g$ increases from
0 to 1. In other words, in contrast to the more usual `linear' electrostatics we now have a far-field monopole
potential that is not only determined by the net charge but in fact {\em also} by its heterogeneity. This is a key
finding, relevant for understanding patchy-particle interactions. The mode coupling has an even stronger effect on
renormalization of $Z_1^*$, for which figure 3(a) and (b) show a pronounced maximum in between the low-$Z_0$ linear
screening regime and the high-$Z_0$ regime in which $Z_1^*$ becomes even {\em vanishingly small} for all $g$'s
considered. Figure 4 shows, for $g=0.4$, that in fact {\em all} $Z_{\ell}^*$ with $\ell\neq 0$ vanish in the limit
of large $Z_0$, while they all show an intermediate regime with finite values even for $\ell=2,4,6$ for which
$\sigma_{\ell}=0$. The underlying physics for non-oppositely charged hemispheres with $\sigma_N>\sigma_S>0$ (that is,
with $0<g<1$) is that both $\sigma_N$ and $\sigma_S$ renormalize, if both are high enough, to the {\em same}
saturated value, giving rise to a pure far-field monopole without multipoles.\\

The idea might emerge that both hemispheres renormalize their charge independent of each other, that is, some of the results could suggest that the renormalized charge density on the colloidal surface is a function which depends only locally on the bare charge density. If that were the case, it would suffice to calculate the renormalized surface charge-density for $\sigma_N$ and $\sigma_S$, as if both were the charge density of a monopole. The cell radius is to be kept unchanged. The thick dotted lines in figure 3(a) and (b) denote the resulting dipole charge for $g=0.8$, which is calculated with the obtained (renormalized) $\sigma^*_N$ and $\sigma^*_S$ by $\frac{3}{4}(\sigma^*_N - \sigma^*_S)$. The correspondence with the solid line is at best reasonable but not perfect. Also the effective monopole is not accurately predicted. This can can be seen from figure 5, where we investigate the cell-size dependence of the the effective monopole for $g=0.4,0.8$ and for $\kappa a = 1,3$. We included data from the full theory (depicted by symbols) and the predicted values by
$2\pi a^2(\sigma_{N}^{*} +\sigma_{S}^{*})$ as a solid line. One can see that the monopole charge is underestimated for a wide range of cell radii, especially for larger cells. Nevertheless, there is qualitative agreement on the increase of the renormalized charge with higher values for $\kappa a$. The difference with the full theory is expected to be the largest for very heterogeneously charged particles. Indeed, we see the largest discrepancy in figure 5(a) and (b) for $g=0.8$, with deviations up to $20$--$25$\% between the results of the present theory for $\sigma_0^*$ and those of the simple approximation $(\sigma_N^*+\sigma_S^*)/2$ discussed above. Apparently, the interactions between the hemispheres do play a role, which in fact can can also be concluded from the induced even multipoles in figure 4. Further research might give more insight into the characteristics of these interactions.
\begin{figure}[!ht]
\centering
\includegraphics[width = 8cm]{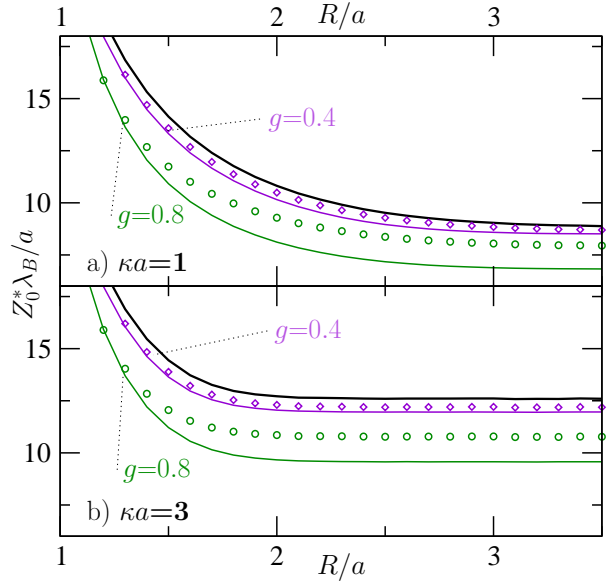}
\caption{Renormalized monopole charge $Z_{0}^*$ for several $g$ of a Janus sphere as a function of the cell size $R$, both for $Z_0 \lambda_B / a = 22.5$, in (a) for $\kappa a=1$ and in (b) for $\kappa a=3$. The thick solid line corresponds to a pure bare monopole, yielding $g=0$. The diamonds ($g=0.4$) and circles ($g=0.8$) show the data from the full theory, using $L=13$. The thinner solid lines denote the average of the renormalized northern and southern charge presumed distributed homogeneously (see text) for $g=0.4$ and $g=0.8$.}
\label{fig:FIGURE5}
\end{figure}

\section{Discussion and Conclusion \label{discussion}}
The newly emerging general picture
is that nonlinear ionic screening of heterogeneously distributed surface charges strongly affects the
far-field symmetry of the potential, and hence also the symmetry of the effective interactions and the
self-assembling structures. The systematics of the present screening theory could be a firm basis to further study these intricate features of heterogeneously charged particles. The category of particles which are described by $g>1$, carrying positive and negative charges on the two hemispheres, are particularly interesting. Because the present method is not very efficient at high values of $g$, we have developed an alternative method to treat these mainly dipolar particles within a mode expansion as well. This will be presented in further work.

It is tempting to model Janus spheres by a Yukawa-segment model \cite{janus2} in which every surface element $dS$
contributes $\sigma^*(x)\lambda_Bs^{-1}\exp(-\kappa s)dS$ to the (dimensionless) potential at a distance $s$. Here the renormalized charge densities on both hemispheres is obtained from the renormalized monopole and dipole charge density via
\begin{eqnarray}
\sigma^*(x) &=&\left\{\begin{array}{ll}\sigma_0^* + \frac{2}{3} \sigma_1^* &x<0;\\\sigma_0^* - \frac{2}{3} \sigma_1^* & x\geq0.\end{array}\right.\
\label{sigmayukawa}
\end{eqnarray}
The dotted curves in figure 1(a) show that agreement with our full calculations is reasonable though not quantitative;
the Yukawa model ignores the ionic hard-core exclusion in the interior of the particle. Therefore, effectively it describes the (dimensionless) potential of a charge configuration in which oppositely charged ions were able to approach the heterogeneously charged surface from two sides, such that this potential is more suppressed compared to the full theory.

Being a point of discussion, we return to the choice of the boundary conditions on the cell's surface (\ref{BCiso}), which we called `isotropic' BC's. This denomination follows from the fact that, considering two randomly oriented neighboring cells, the cell-surface potentials of two cells should match on the spot where they touch, giving rise to a constant cell-surface potential. Nevertheless the choice of BC's is not unique. We can also supply the system with `nematic' boundary conditions, corresponding to the situation that all cells are perfectly aligned such that cells only touch on opposite spots. On these spots the electrostatic potential should match, and we can also demand continuity of the electric field. The BC's then become
\begin{eqnarray}
\Phi(R,\theta,\varphi) &=& \Phi(R,\pi-\theta,\pi+\varphi), \nonumber \\
\Phi^{'}(R,\theta,\varphi) &=& - \Phi^{'}(R,\pi-\theta,\pi+\varphi). \label{BCnem}
\end{eqnarray}
In fact one can even interpolate between `isotropic' and `nematic' BC's by introducing an orientation distribution function \cite{eelco}. In this article, we assume a system in which the cell boundary is best described by isotropic BC's, given by (\ref{BCiso}). However, for the parameters used in this article, it turns out that this particular choice for the BC's did not noticeably affect the values of the renormalized charges. We do not see a significant change by turning to nematic BC's (\ref{BCnem}). This insensitivity to the choice of BC's is due to the fact that the nonlinear behaviour is an effect which takes place close to the colloidal surface, where these BC's have the least influence on the electrostatic potential. Furthermore, the only multipoles which are directly affected by the particular choice of BC's are the nonzero($\ell>0$) even multipoles, which are small for Janus particles. We therefore think that the obtained values for the renormalized multipoles can be applied in a model to describe the behaviour of a many-body system within linear theory, no matter what the orientations of the surrounding colloids are. Since the monopole and multipole potentials decay for large $r$ equally fast as $\exp(-\bar{\kappa}r)/r$ \cite{ramirez}, where $\bar{\kappa}^{-1}$ is the decay length, the renormalized multipole charges are expected to contribute in dense as well as dilute systems.
\\

In summary, we have developed a systematic framework for nonlinear ionic screening of heterogeneously charged
spheres. The scheme allows for an explicit far-field analysis that generalizes charge renormalization from the
well-studied homogeneous case (pure monopole) \cite{alexander} to the heterogeneous case (dipoles, quadrupoles,
etc. and their nonlinear couplings). Application to charged Janus spheres shows (i) a 40\% reduction of the
effective monopole for $g=1$ (charged and uncharged hemisphere) compared to $g=0$ (homogeneously charged sphere),
(ii) a mode-coupling induced far-field effective quadrupole component without an actual surface quadrupole, (iii)
a pure far-field monopole with vanishing higher-order multipoles in the saturated high-charge limit, and (iv) no
quantitative agreement with a simple Yukawa-segment model based on renormalized multipoles. Our study opens the
way to systematic microscopic calculations of effective electrostatic interactions between Janus (and other
patchy) particles. In addition, our analyses also reveal non-vanishing electric fields {\em inside}
heterogeneously charged particles, which could couple to interior dipoles and affect (anisotropic) mutual Van der
Waals forces. Given that the presently introduced expansion technique can be generalized to other geometries (for example, patterned planar surfaces or ellipsoidal patchy colloids), our technique and findings are directly relevant for
gaining microscopic understanding of effective interactions and ultimately phase behaviour of a large class of
dispersions of patchy or patterned nanoparticles, colloids, or proteins \cite{glotzer}.

This work is part of the research program of the Stichting voor Fundamenteel Onderzoek der Materie (FOM), which is financially supported by the Nederlandse Organisatie voor Wetenschappelijk Onderzoek (NWO). Financial support from NWO-CW-ECHO, NWO-VICI, and FOM-SFB-TR6 is acknowledged.

\end{document}